\documentstyle[11pt]{article}
\begin{document}
\begin{center}
{\bf Stellar Population of Ellipticals in Different Environments:\\
Near-infrared Spectroscopic Observations}
\vskip 3truecm

{\bf P. A. James$^1$ and B. Mobasher$^2$}
\vskip 2truecm

$^1$ Astrophysics Research Institute, Liverpool John Moores 
University,\\ 
Twelve Quays House, Egerton Wharf, Birkenhead L41 1LD.\\
$^2$ Astrophysics Group, Blackett Laboratory, Imperial
College,\\ 
Prince Consort Road, London SW7 2BZ.
\vskip 1 truecm

{\it Accepted for publication in the Monthly Notices of the Royal 
Astronomical Society}

\vskip 1 truecm

\end{center}

\centerline {\bf ABSTRACT}

\noindent
Near-infrared spectra of 50 elliptical galaxies in the Pisces, A2199 \&
A2634 clusters, and in the general field, have been obtained. The
strength of the CO (2.3 $\mu m$) absorption feature in these galaxies is
used to explore the presence of an intermediate-age population (e.g.
Asymptotic Giant Branch stars) in ellipticals in different environments.
We find the strongest evidence for such a population comes from
ellipticals in groups of a few members, which we interpret as the result
of recent minor merging of these galaxies with later type galaxies.
Field galaxies from very isolated environments, on the other hand, show
no evidence for young or intermediate-age stars as revealed by H$\beta$
and CO absorptions, and appear to form a very uniform, old population
with very little scatter in metallicity and star formation history.


\vfil\break
 
\noindent{\bf 1 INTRODUCTION} 

\noindent
Elliptical galaxies are conventionally assumed to have stellar
populations dominated by old stars, formed in a single burst some
12--16~Gyr ago.  However, the degree to which this assumption is violated
in observed systems is very uncertain.  Indeed, some of the studies regarded
as underpinning this view express distinct reservations on the accuracy
of this picture. For example, Tinsley \& Gunn (1976) point out that
their modelling of optical and near-infrared colour and line indices of
ellipticals cannot rule out the existence of `a considerable population
of stars born after the rapid initial burst' (see also Bruzual 1983).
However, more recent studies show that less than 10 per cent of the stars in
present-day ellipticals are likely to have been formed in the last 5~Gyr,
imposing strong constraints on the time and duration of the last episode
of star formation in ellipticals (Bower, Lucey \& Ellis 1992). Moreover,
using UV-optical colours of ellipticals in clusters at $z\simeq$0.5,
Ellis et al. (1997) confirm that star formation in most cluster
ellipticals was essentially completed by $z\simeq$3.

However, one outstanding question is whether field ellipticals, which are
currently isolated or surrounded by at most a small number of neighbours,
have the same star formation history as their cluster counterparts.
Indeed, several observational studies have found strong evidence that
they do not, with the star formation extending to significantly more
recent times in the case of field ellipticals.  For example, Larson,
Tinsley \& Caldwell (1980) found that bright field ellipticals are, on
average, bluer than cluster ellipticals of the same luminosity, implying
more recent star formation in the former. Furthermore, O'Connell (1980)
showed that major star formation episodes in the nearby blue elliptical
M~32 continued until $\sim$5~Gyr ago while, using optical spectroscopy,
Bica \& Alloin (1987) concluded that field ellipticals and lenticulars
contain an intermediate-age component, not present in cluster
ellipticals. These results were further confirmed by Bower et al. (1990)
and Rose et al. (1994) who used a range of optical spectral indicators to
infer that `a substantial intermediate-age population is present in the
early-type galaxies in low-density environments, a population that is
considerably reduced or altogether lacking in the early-type galaxies in
dense clusters'. Similarly, Schweizer \& Seitzer (1992) found that some
field ellipticals have blue UV-optical colours and strong H$\beta$
absorption lines, indicating star formation within the last 3--5~Gyr.
Also, the range in H$\beta$ strength observed in field ellipticals was
interpreted by Gonzalez (1993) as evidence for an inhomogeneous stellar
population. Finally, Kauffmann (1996) showed via simulations of the star
formation and merging history of galaxies that star formation can
naturally be expected to continue to more recent epochs in low-density
environments. An implication of these results, explored by de Carvalho \&
Djorgovski (1992) and Guzm\'an \& Lucey (1993), is that the younger stellar
populations in ellipticals in low-density environments results in an
increase in scatter about, and offsets from, the Fundamental Plane of
elliptical galaxies (Dressler et al. 1987, Djorgovski \& Davis 1987).
This has strong implications regarding formation of ellipticals and
determination of galaxy distances using the Fundamental Plane.

However, there has recently been an important dissenting opinion, in the
work of Silva \& Bothun (1998), who study the near-IR colours of a
sample of field elliptical galaxies with signs of recent disturbance
(i.e. possible post-mergers). They concluded that $<$10--15 per cent of the
stellar mass in these systems is in the form of intermediate-age stars,
with ages 1--3~Gyr. The fraction of light contributed by such a
population can be somewhat larger, but is still $<$20--30 per cent. This is a
surprising result, because this sample contains just those field galaxies
which would be expected to show the strongest signatures of recent star
formation.

To address these questions, we carried out a spectroscopic study of
elliptical galaxies in different environments, using near-infrared
measurements of the 2.3~$\mu$m CO 2--0 photospheric absorption feature to
determine the contribution from the intermediate-age stellar population
to the light of these galaxies (Mobasher \& James 1996; henceforth paper
I).  We found marginally-significant evidence for deeper CO absorptions,
and hence larger intermediate-age populations, in the field galaxies.
However, the small sample size (9 field and 12 cluster galaxies) limits
the statistical significance of these results.  The present paper extends
the sample to 50 ellipticals in total (20 field and 30 cluster) and looks
in more detail at optimal techniques for extracting information on
stellar populations from CO measurements. The near-IR spectral data
presented here are for ellipticals in the Abell~2199 and Abell~2634
clusters, and for field galaxies from Faber et al. (1989).

Recent observations and data reduction are outlined in section 2. Section
3 presents results regarding the correlation between star formation
history and environment. Section 4 explores the dependence of CO index on
the physical parameters in galaxies and discusses some of the relevant
peculiarities of individual galaxies.  Section 5 contains a comparison of
these results with previous studies, and section 6 summarises the main
conclusions.

A Hubble constant of 75~km~s$^{-1}$~Mpc$^{-1}$ is assumed in this paper.

\noindent{\bf 2 OBSERVATIONS AND DATA REDUCTION}

\noindent
The observations presented here were carried out using the United Kingdom
Infrared Telescope (UKIRT) on the 4 nights 1996 August 2--5.  The
instrument used was the long-slit near-IR spectrometer CGS4, with the
150~line~mm$^{-1}$ grating and the short-focal-length (150~mm) camera.  The
2-pixel-wide slit was chosen, corresponding to a projected width on the
sky of 2.4~arcsec.  The wavelength range in this configuration
in first order was 0.33~$\mu$m, centred on the start of the CO absorption band
at $\sim$2.33~$\mu$m for galaxies with recession velocities of a few
thousand km~s$^{-1}$ (the present sample has redshifts from
1500--12000~km~s$^{-1}$).  The effective resolution, 
including the effective
degradation caused by the wide slit, was about 900, and the array was
moved by 1 pixel between integrations to enable bad pixel replacement in
the final spectra.

For each observation, the galaxy was centred in the slit by maximising
the IR signal, using an automatic peak-up facility.  Total integration
times were between 40 minutes and an hour per galaxy, depending on their
central surface brightness.  During this time, the galaxy was slid up and
down the slit at one minute intervals by 22~arcsec, giving two
offset spectra which were subtracted to remove most of the sky emission.
Stars of spectral types A0--A6, suitable for monitoring telluric
absorption, were observed in the same way
before and after each galaxy, with airmasses matching those of the galaxy
observations as closely as possible (root-mean-square difference
$<$0.1~A.M.). Flat fields and argon arc spectra were taken using the CGS4
calibration lamps.

The spectra were reduced using the FIGARO package in the Starlink
environment, as outlined in paper I.  However, there was an initial
complication, resulting from the CGS4 slit rotation mechanism having
jammed at the time that these observations were made.  This resulted in
the slit being substantially misaligned with the columns of the array,
which made sky-line subtraction and spectral extraction difficult.  The
problem was overcome by using the FIGARO routine SDIST to fit the
orientations of arc lines in spectra taken during each of the 4 nights.
This resulted in a correction which was applied to each spectrum taken
using the CDIST routine.  This worked very successfully, which was
demonstrated by applying the correction obtained from one arc spectrum to
another taken on the same night.  Arc lines in the corrected spectra
were perfectly aligned with the array columns, and the only side-effect
was a slight loss of wavelength range as the ends of the corrected
spectra subsequently had to be trimmed.  This had no impact on the
present programme, since the available wavelength range was in excess of
that needed.  The galaxy flux was then extracted from $\sim$5 pixels (i.e.
6~arcsec.) along the slit.

The result of the data reduction was a one-dimensional,
wavelength-calibrated galaxy spectrum, which had been divided by an
A-star spectrum to remove the effects of atmospheric absorptions. This
was converted into a normalised, rectified spectrum by fitting a
power-law to featureless sections of the continuum, and dividing the
whole spectrum by this power-law, extrapolated over the full wavelength
range (see Doyon, Joseph \& Wright 1994 for a discussion of this
procedure).  This fitting process made use of code kindly written by Dr
C. Davenhall under the Starlink QUICK facility. The resulting rectified
spectrum then has a flat continuum level of unity across the whole
wavelength range, simplifying the calculation of equivalent widths and
spectral indices of the spectral features.  The spectra were wavelength
calibrated using the argon arc spectra, and redshift-corrected on the
basis of their catalogued recession velocities.

To quantify the depth of the 2.3~$\mu$m CO absorption features, several
different methods have been proposed. For example, Doyon et al. (1994)
use a spectroscopic CO index, defined by $$ CO_{sp} = -2.5 log <R_{2.36}>
$$ where $<R_{2.36}>$ is the average value of the rectified spectrum
between 2.31 and 2.4~$\mu$m in the galaxy rest frame.  Doyon et al.
(1994) give conversions between this index and the photometric CO index
used by earlier studies based on narrow-band filter observations, and
also calibrate the index against effective temperature for dwarf, giant
and supergiant stars.  This was the definition we used to quantify CO
depth in paper I.  However, Puxley, Doyon \& Ward (1997; PDW) have
recently proposed a new definition, which they claim to be the most
powerful discriminant between different stellar populations in galaxies.
Considering various options, they conclude that the optimal wavelength
range is 2.2931--2.32~$\mu$m, a substantially narrower range than for
$CO_{sp}$, and they express this as an equivalent width in nm, rather than as
an index in magnitudes. Whilst they adopt this measure for astrophysical
reasons, there are practical advantages resulting from the narrower
wavelength range. Errors resulting from the uncertainty in the power-law
fit to the continuum are substantially reduced, because the degree of
extrapolation needed is much smaller.  Also, the new definition enables
higher-redshift objects to be observed without the spectra becoming
unduly noisy as the bandpass of interest moves to the end of the K
window.  On the Mauna Kea site, the window is typically usable to about
2.5~$\mu$m, which corresponds to a limit of only 12,500~km~s$^{-1}$ in
recession velocity before the end of the $CO_{sp}$ range is lost,
compared to 23,000~km~s$^{-1}$ for the range advocated by PDW.

However, there are arguments against going to the still shorter
wavelength range used by, for example, Kleinmann \& Hall (1986) and
Origlia, Moorwood \& Oliva (1993). Their EW measurements were found to be
sensitive to velocity-dispersion smoothing, whereas the PDW find their EW
to be completely unaffected (we apply no velocity dispersion corrections
to the EW in the present paper).  In addition, shorter baselines result
in reduced signal-to-noise, which would be a significant problem for the
fainter galaxies here. Thus, we calculate both the CO EW defined by PDW,
and the $CO_{sp}$ index for comparison with paper I and the calibrations
of Doyon et al. (1994).  Whilst the prescription given by Doyon et al.
(1994) and PDW are simple and, it is to be hoped, unambiguous, the
resulting indices and equivalent widths should be regarded as
instrumental measures. It would be useful to obtain repeat measurements
of the present sample with other telescopes and instruments to check for
possible systematic differences, but this has not yet been done.

In calculating the errors in the CO estimates, a number of random and
systematic sources of error were taken into account.  Random errors
consist of both photon counting statistics and weak unresolved spectral
features, which effectively introduce similar errors. These were
estimated from the standard deviation of the `featureless' continuum used
in the power-law fitting, after division by the power-law.  This
constitutes the dominant error in the CO EW values, due to the relatively
small spectral range on the CO bandhead, whilst the random error on the
continuum determination is small because most of the continuum between
2.15 and 2.28~$\mu$m can be used. This gives a 1--$\sigma$ error on the
CO EW estimated at 0.2~nm. Determination of the continuum slope from the
power-law fitting also contributes to the error, and was estimated by
varying the wavelength ranges used in the power-law fitting, and by using
different packages and algorithms for the fitting, with errors of 0.2~nm
allocated to this cause.  Finally, errors due to wavelength calibration
and redshift uncertainty were estimated by shifting the spectral bandpass
by an amount equivalent to $\pm$400~km~s$^{-1}$, and an error of 0.1~nm
allocated to this for all measurements. Adding these three sources of
error in quadrature, since they are most plausibly uncorrelated with one
another, overall errors of 0.3~nm were assigned to the CO EW values.

The CO measurements for the sample of 29 elliptical galaxies in this
study (both in field and clusters) are listed in Table 1. This also
contains the 21 galaxies from paper I for which new CO measurements have
been determined following the prescription of PDW.  Column 1 gives
the galaxy name.  For the cluster galaxies without NGC, UGC or IC
numbers, the designator given is either from Butcher \& Oemler (1985) or
Lucey et al. (1997), and these galaxies are listed in the format BO$nnn$
or L$nnn$ respectively.  Column 2 gives the heliocentric recession
velocity in km~s$^{-1}$, column 3 $CO_{sp}$, and column 4 the CO EW as
defined by PDW, in nm.  Column 5 gives the H$\beta$ equivalent width from
Trager et al. (1998), in units of 0.1~nm.  Column 6 gives the total B-band
absolute magnitude, calculated from B$_T$ values in the NASA/IPAC
Extragalactic Database (NED) and the heliocentric redshift. Column 7
gives the log of velocity dispersion in km~s$^{-1}$ and column 8 the
$Mg_2$ index in magnitudes, both of which were taken from Faber et al.
(1989) or Lucey et al. (1997). Column 9 gives the cluster or group
membership where HG and GH denote groups from Huchra \& Geller (1982) and
Geller \& Huchra (1983) respectively, and `Isol' denotes galaxies which
pass our isolation criterion, which we describe in section 3.

As a check on the overall validity of our data and reduction methods, we
calculated the stellar type of stars giving rise to the near-IR light in
these galaxies. This involved converting the CO EW values to CO$_{sp}$
using the equation given by PDW, then using the CO$_{sp}$--effective
temperature calibration given by Doyon et al. (1994), and finally the
effective temperature--spectral type conversion from Allen (1976). We find
that the CO absorption strengths for most galaxies are consistent with
those of K giants, with a range equivalent to stars of spectral types
K3III--K8III. Photometric studies (e.g. Frogel et al. 1978) tend to show
that the near-IR light of galaxies is equivalent to that of late K or
early M giants, reasonably consistent with our findings.

\noindent{\bf 3 ENVIRONMENTAL DEPENDENCE OF CO DEPTH}

\noindent
The distribution of CO EWs for the 50 elliptical galaxies in this
study is shown in Fig. 1, where the cluster and field galaxy
distributions are shown by solid and dashed lines respectively. The field
galaxy distribution is also shaded for clarity. The field galaxies appear
to cover a broader range of CO EWs compared to their counterparts in
clusters. The field ellipticals also show a pronounced bimodal
distribution. Overall, the mean CO strengths for the field and cluster
galaxies are similar, corresponding to 3.28$\pm$0.12~nm and
3.29$\pm$0.06~nm respectively, with the errors being the standard
deviations from the mean. However, the apparent difference in the
distributions is moderately significant, with a K-S test determining that
there is only an 8 per cent chance that the two distributions were drawn
from the same parent population. In paper I, the field galaxy
distribution was simply displaced towards stronger CO indices (or larger
EW values) compared with that for clusters.

However, there is a potential problem, in that the two peaks found in the
CO EW distribution for field galaxies correlate strongly, though not
perfectly, with the observing run (there is no such correlation for cluster
galaxies). This is in the sense that most of the galaxies in the high EW
peak centred on $\sim$3.8~nm were observed in the 1994 run described in
paper I, whereas the peak centred on 2.7~nm is comprised of galaxies
observed in the 1996 run.  The difference is significant, with the field
galaxies in the 1994 run having a mean CO EW of 3.82$\pm$0.06~nm, whereas
those observed in the 1996 run have a mean of 2.88$\pm$0.09~nm.  This
immediately raises the question of whether the offset is a systematic,
due to some change in the instrument or the data reduction procedures.
We take this possibility extremely seriously, and in the following
discussion examine all the possibilities to explore the reality of this
effect. We emphasize here that the CO EW values for the cluster galaxies
observed in 1994 had an average value of 3.34$\pm$0.09~nm, compared with
3.25$\pm$0.08~nm for the cluster galaxies observed in 1996.  Thus the
cluster galaxy distributions show no offset between the two runs, as is
confirmed by a K-S test, and any systematic would therefore have
preferentially to affect the field galaxies.

We first checked whether the stars used to remove atmospheric
absorptions have had any effect on the measured CO EW values.
However, all of the stars were drawn from a very small range of
spectral types, with no chance of any intrinsic CO absorption.  They were
bright enough that the chance of misidentification is very slight, and in
any case, no correlation was found between the derived CO EW and the
star used.  There are many cases where the same
star was used for galaxies which were found to have CO EW values
drawn from opposite extremes of the measured range, and conversely using
different stars for a given galaxy spectrum was found to have no
effect on derived EW values within the errors.  Airmass differences
between the galaxy and star observations were also found to have no
measurable effect on the CO EWs.

We checked for wavelength calibration differences between the two
runs, but no systematic difference was found.  The CO EW definition
adopted by PDW is, in any case, only weakly sensitive to wavelength errors,
and any shift large enough to give the observed offset between the two field
galaxy groups in Fig. 1 would have been immediately apparent in the
spectra.  The same argument also excludes redshift errors as the source
of this effect, with shifts of $\pm$400~km~s$^{-1}$ only changing the
derived EW by $\pm$0.1~nm in the most extreme cases.

The instrument configuration was quite different between the two runs,
and indeed a different detector array had been installed.  The main
consequences of this were a longer wavelength range for the 1996 run,
giving a longer continuum baseline, and a somewhat coarser wavelength
resolution for the 1994 observations.  This gives rise to a number of
possible effects on the derived CO values.  One concerns the power-law
fitting and extrapolation, which is better constrained for the more
recent data, and the difference in spectral resolution could also
potentially cause an effect due to rounding error in the wavelength range
over which the equivalent width was calculated.  To check for any
systematic effects on the derived CO strengths, we re-reduced the more
recent spectra, reducing the wavelength range and rebinning into coarser
pixels in the wavelength direction, to match the wavelength range and
resolution of the 1994 data.  CO strengths were then derived from these
simulated spectra, and were found to differ by up to 0.25~nm from values
originally derived. In every case, most of the change was due to 
fitting the power-law continuum over a smaller
wavelength range, as was demonstrated by producing spectra where the
wavelength range and the resolution were changed individually.  However,
even the largest differences found by changing both wavelength range and
resolution are much smaller (by a factor of at least 4) compared with the
offset between the two field galaxy groups shown in Fig. 1.

Finally, the
Galactic latitudes of the field galaxies were checked to explore if
extinction effects might be significant.  However, both the high-EW and
low-EW groups have $|b|$ distributions which are approximately uniform
over the range 20$^{\circ}$ and 60$^{\circ}$, with no galaxies outside
these limits.

Thus we have checked all the potential systematic effects of which we are
aware, and none contributes significantly to the offset between the 
two groups of field galaxy CO strengths in Fig. 1.  We therefore conclude
that this represents a real difference between the galaxies, and continue
now to discuss astrophysical explanations for this.

When selecting field galaxies for the 1994 run, we based our choice
purely on information tabulated by Faber et al. (1989), checking only
that `field' galaxies were not members of major clusters. However, for
the 1996 run, we selected field galaxies from highly isolated
environments, thus maximising the environmental difference between
field and cluster samples.  Thus there is a strong correlation between
degree of isolation and observing run, which we now propose explains the
offset in field galaxy properties. Although these are all nominally field
galaxies, there is in fact quite a wide range of environments sampled by
these 19 galaxies, ranging from complete isolation up to membership of
groups with $\sim$10 bright members.  The criterion we adopt for complete
isolation is that there be no companions within $\sim$5 magnitudes in
apparent brightness, within a projected radius of 500~kpc.  Of the 19
field galaxies, 6 meet this criterion (IC~5157, NGC~6020, NGC~6127,
NGC~7391, NGC~7785 and ESO462--G015).  At the other extreme, 4 galaxies
are found in groups rich enough to have been identified by Huchra \&
Geller (1982) or Geller \& Huchra (1983), who employed an algorithm
based on position and redshift information only, to identify significant
groupings in the CfA redshift survey and another magnitude-limited,
all-sky sample. These galaxies are identified in Table 1.  In addition,
NGC~1600 is generally considered to lie at the center of a group of at
least 10 members.  The other 8 members of our field subset have at least
one apparent companion but are fairly isolated, lying in groups with at
most a few members.

It is interesting to note that all 6 completely isolated galaxies lie in
the low--CO peak while the 5 galaxies in moderately rich groups lie in
the high--CO peak in Fig. 1.  The remaining `field' galaxies, with a
small number of apparent companions, are split between the two peaks. An
interpretation of this result will be discussed in section 5.

It is also of interest to investigate the possible correlation of CO EW
with environment for the cluster galaxies. However, Fig. 2 shows 
no correlation between CO EW and projected distance of galaxies
from cluster centres.

\noindent{\bf 4 SPECTROSCOPIC PROPERTIES OF ELLIPTICALS}

\noindent
In this section, we briefly look at correlations between the
spectroscopic CO EWs and other physical parameters of ellipticals in our
sample. 

A weak dependence is found between the CO EW and the metallicity of
ellipticals, with the latter being measured by the $Mg_2$ index (Fig. 3).
There is the hint of a correlation in the expected sense, but this is
formally only at the 1--$\sigma$ level. This is unsurprising given that
previous studies have shown the CO index to be a poor metallicity
indicator in old and metal-rich stellar populations. For example, Frogel
et al. (1978) found no gradient in the photometric CO index with radius,
even though galaxies have strong metallicity gradients as shown by
optical indicators. There is a correlation between $Mg_2$ and absolute B
magnitude, in the sense that more luminous galaxies have higher
metallicity (Fig. 4). However, this does not reflect in any measureable
correlation between CO EW and absolute B magnitudes, as shown in Fig. 5,
which also shows that the scatter in CO EW is constant with galaxy
luminosity. Fig. 6 shows CO EW plotted against H$\beta$ equivalent width,
for the 19 galaxies in the present sample which were observed in the
optical by Trager et al. (1998).  Since H$\beta$ is strong in stellar
populations with relatively recent star formation, it is reassuring to
see a trend, however weak, in the expected sense in this plot, but the formal
significance of the correlation is low due to the small number of
objects for which data are available, and due to the different ages of
populations probed by the two indicators.

Some of the individual galaxies have known peculiarities which might
affect their star formation histories and hence the measured CO EWs. 
For example, 
NGC~1052 hosts a liner-type nucleus with broad H$\alpha$ emission and
optical--near-IR line ratios inconsistent with stellar excitation
(Alonso-Herrero et al. 1997), and is thus concluded to have a central
nonstellar ionizing source.  NGC~6051 is the radio source 4C+24.36 (Olsen
1970). NGC~6137 is identified as a head-tail radio source by Ekers
(1978). NGC~6166, a multiple-nucleus cD galaxy in Abell 2199, is a strong
radio source (3C338).  Carollo et al. (1997) note that HST WFPC2 images
of NGC~7626 show a warped, symmetrical dust lane crossing the centre, and
this galaxy also has strong radio jets (Jenkins 1982).  The
compilation of radio surveys by Calvani, Fasano \& Franceschini (1989)
contains radio observations for 12 of the present sample; of these, 5 are
detections (NGC~1052, NGC~1600, NGC~7385, NGC~7626 \& NGC~7785) while the
remainder (IC~5157, NGC~380, NGC~384, NGC~410, NGC~6020, NGC~7454 \&
NGC~7562) are all upper limits. Since any links between star formation
and nuclear activity are highly uncertain at present, we make no comment
on these data, other than to note that no strong correlation of radio
properties with CO EW is evident.

\noindent{\bf 5 DISCUSSION}

\noindent
The main issue confronted in this study is the effect of environment on
the stellar population in elliptical galaxies (i.e. the epoch of star
formation). Using the CO measurements for field and cluster ellipticals,
we find no overall evidence for a stronger CO absorption feature in field
galaxies compared to those in clusters, contrary to what was found in
paper I. However, we find a clear difference between the CO strengths of
ellipticals in small groups, with a few companions, and those which are
truly isolated. Galaxies located in groups have CO EW values larger than
either the isolated field subset, or than the cluster galaxies.
Therefore, a simple monotonic correlation between star formation history
and local galaxy number density seems untenable. This is the most
important result from the present study.

However, it is unlikely to be a coincidence that the ellipticals appearing to
contain young stars are preferentially located in environments most
conducive to merging, due to the lower velocity dispersion of the groups
compared to rich clusters, and indeed to merging with late-type
companions, which are far more common in groups than rich clusters. Thus
we propose, along with Ellis et al. (1997), that {\em all} ellipticals
formed most of their stars at an early epoch, and those containing
younger stars have acquired them recently through minor mergers with
gas-rich galaxies. This is also suggested by Silva \& Bothun (1998) in a
study using near-infrared photometry of ellipticals which show signs of
interaction/mergers. It is, however, likely that our spectroscopy, which
only samples the central few arcsec, is more sensitive to such
a component if the merging is strongly dissipative, with the accreted
material rapidly sinking to the centre. Thus, our data should be more
sensitive to nuclear starbursts than would be the global colours measured
by Silva \& Bothun (1998).

The observed offset between the isolated field subsample and the main
peak of cluster galaxies in Fig. 1 is not a metallicity effect since they
have similar mean $Mg_2$ indices (which are measures of metallicity),
corresponding to 0.304$\pm$0.004~mag. (for the `isolated' ellipticals)
and 0.295$\pm$0.005~mag. (for the cluster ellipticals). Therefore, given
the weak correlation between $Mg_2$ and CO EW (Fig. 2), the offset
between isolated and cluster galaxies must have some other cause.
However, there is a striking difference between the {\em range} of $Mg_2$
indices for the different subsamples.  The standard deviation of the
$Mg_2$ indices for the isolated galaxy subsample is only 0.01~mag, c.f.
0.03~mag for the cluster galaxies and 0.04~mag for the group galaxies.
While we must beware small number statistics (only 6 of the isolated
galaxies have $Mg_2$ values), this indicates that the isolated
ellipticals form a very homogeneous group of galaxies, a point which is
also indicated by the extreme narrowness of the corresponding peak in CO
EW values shown in Fig. 1.  Conversely, the large range in $Mg_2$ for the
group galaxies is easily understood if these are prone to significant
merging and accretion activity.

There have been several studies claiming to have found intermediate-age
populations in field galaxies (e.g. Bica \& Alloin 1987, Bower et al.
1990, Schweizer \& Seitzer 1992, Rose 1994).  It is therefore of interest
to see whether these are consistent with the interpretation presented
above. Bica \& Alloin (1987) find 5 individual galaxies which show clear
evidence of intermediate-age or young stellar populations.  Of these, 4
(NGC~2865, NGC~5018, NGC~5061 \& NGC~5102) are listed by Faber et al.
(1989) as being members of groups with group velocity dispersions between
120 and 430~km~s$^{-1}$.  The only exception is NGC~4382, an interacting
S0 on the outskirts of the Virgo cluster.  Thus, all the 5 galaxies are
located in environments conducive to merging activity. The same is true
for the ellipticals studied by Rose et al. (1994), where all of the
field ellipticals lie in the outer regions of the Virgo cluster, or in
groups identified by Faber et al. (1989), with group velocity dispersions
between 65 and 210~km~s$^{-1}$. Since the majority of ellipticals lie in
groups or clusters, this can hardly be taken as a striking confirmation
of our model, but it is encouraging that there are no discrepant objects
in these two independent studies.

Further observations are clearly required to test our present
interpretation.  CO spectroscopy should be obtained for field galaxies
claimed by other studies (e.g. Bica \& Alloin 1987; Rose et al. 1994) to
have undergone recent star formation, and for the apparent post-merger
objects of Silva \& Bothun (1998).  It is also important to extend the
present sample of very isolated galaxies to confirm the very homogeneous
old populations we have found in the present paper.  Our
present interpretation in terms of recent merging predicts a correlation
between CO absorption strength and the velocity dispersion of the group
or cluster in which the galaxy resides; it will be of interest to see
whether such a correlation is confirmed by future observations of galaxies
in a range of environments from the core of the Coma cluster to the field.

In order to calibrate the observed differences in CO strengths in terms
of starburst ages and strengths, reliable evolutionary synthesis models
are required.  This is an area which is yet to be fully developed in
terms of near-IR spectroscopic parameters, but some preliminary work has
been done.  Buzzoni (1995) tabulates CO indices for simulated stellar
populations with ages between 4 and 15~Gyr, and in general predicts
strengths similar to those in the isolated field galaxies in the present
sample.  For example, taking a Salpeter IMF and solar metallicities, the
CO indices predicted by Buzzoni (1995) convert to equivalent widths of
2.4--3.0~nm, using the conversion formulae given by Doyon et al. (1994)
and PDW.  This agreement is encouraging, but it should be noted that
these models cannot explain the strength of CO absorption observed in the
`group' ellipticals.  Even if there is an overall offset between the
model predictions and observations, it is interesting to note that
Buzzoni et al. (1995) find only $\sim$10\% change in CO absorption
strength between 4~Gyr and 15~Gyr populations (the extremes in the
modelled ages), for any reasonable IMF and metallicity.  The $\sim$30\%
difference observed between the `isolated' and `group' galaxies thus
requires either a population of stars significantly younger than 4~Gyr in
the `group' galaxies, a combination of age and metallicity effects (but
see the discussion earlier in this section) or may point to an additional
stellar component which is not accurately reproduced in the existing
models.  This latter is quite possible, given the complexity of evolution
of giant and supergiant stars.

\noindent{\bf 6 CONCLUSIONS}

\noindent
Contrary to the main conclusion of paper I, we do not find evidence for
an overall offset in CO absorption strength, between field and cluster
ellipticals. However, there is a bimodal distribution in this parameter
for field galaxies only, with the two distributions directly correlated
with the degree of isolation of galaxies. Specifically, very isolated
ellipticals appear to form a very homogeneous population, with no sign of
recent star formation and a very small range of metallicity as revealed
by their $Mg_2$ absorption feature.  On the other hand, ellipticals in
groups frequently show evidence for intermediate-age stellar populations
and have a wide range in metallicity.  Ellipticals in rich clusters have
intermediate properties in both parameters.

We interpret the observed differences in terms of recent minor mergers,
which are most likely to occur in the moderate density environment of
small groups consisting of only a few members.  Dissipative mergers with
gas-rich galaxies could then introduce a significant population of
younger stars to the central regions of galaxies in these groups, giving
the stronger CO absorptions we find.  If this is the case, then our data
are consistent with ellipticals in all environments being essentially
old, in agreement with other recent studies.

Further work is clearly needed to extend the size of the isolated and
group samples, to test further whether the somewhat {\em a posteriori}
division of the field galaxies is actually justified.  In addition, the
whole area of near-IR spectroscopy of galaxies is ripe for detailed
stellar spectral synthesis modelling, so that data of the type we present
here can be fully understood.

\noindent
{\bf ACKNOWLEDGMENTS}

\noindent
We thank the anonymous referee for several useful recommendations which
significantly improved the content and presentation of this paper.
PJ thanks Doug Burke for useful suggestions. This research has made use
of the NASA/IPAC Extragalactic Database (NED) which is operated by the
Jet Propulsion Laboratory, California Institute of Technology, under
contract with the National Aeronautics and Space Administration.
The United Kingdom Infrared Telescope is operated by the Joint Astronomy
Centre on behalf of the U.K. Particle Physics and Astronomy Research Council.

\newpage

\noindent {\bf REFERENCES}

\noindent Allen, C. W., 1976, {\it Astrophysical Quantities (3rd ed.)},
Athlone Press, London\hfil\break
\noindent Alonso-Herrero A., Rieke M. J., Rieke G. H., Ruiz M., 1997,
ApJ, 482, 747\hfil\break
\noindent Bica E., Alloin D., 1987, A\&A, 181, 270\hfil\break
\noindent Bower R. G., Ellis R. S., Rose J. A., Sharples R. M., 1990, AJ,
99, 530\hfil\break
\noindent Bower R. G., Lucey J. R., Ellis R. S., 1992, MNRAS, 254,
601\hfil\break
\noindent Bruzual G., 1983, ApJ, 273, 105\hfil\break
\noindent Butcher H. R., Oemler A., 1985, ApJS, 65, 665\hfil\break
\noindent Buzzoni A., 1995, ApJS, 98, 69\hfil\break
\noindent Calvani M., Fasano G., Franceschini A., 1989, AJ, 97,
1319\hfil\break
\noindent Carollo C. M., Franx M., Illingworth G. D., Forbes D. A., 1997,
ApJ, 481, 710\hfil\break
\noindent de Carvalho R. R., Djorgovski S., 1992, ApJ, 389, L49\hfil\break
\noindent Djorgovski, S., Davis, M., 1987, ApJ., 313, 59\hfil\break
\noindent Doyon R., Joseph R. D., Wright G. S., 1994, ApJ, 421, 101\hfil\break
\noindent Dressler, A., Lynden-Bell, D., Burstein, D., Davies, R.L.,
Faber, S.M., Terlevich, R.J., Wegner, G., 1987, ApJ, 313, 42\hfil\break
\noindent Ekers R. D., 1978, A\&A, 69, 253\hfil\break
\noindent Ellis R. S., Smail I., Dressler A., Couch W. J., Oemler A.,
Butcher H., Sharples R.M.,  1997, ApJ, 483, 582\hfil\break
\noindent Faber S. M., Wegner G., Burstein D., Davies R. L., Dressler
A., Lynden-Bell D., Terlevich R. J., 1989, ApJS, 69, 763.\hfil\break
\noindent Frogel J. A., Persson S. E., Aaronson M., Matthews K., 1978,
ApJ, 220, 75\hfil\break
\noindent Geller M. J., Huchra J. P., 1983, ApJS, 52, 61 (GH)\hfil\break
\noindent Gonzalez J. J., 1993, Ph.D. thesis, Univ. California, Santa
Cruz\hfil\break
\noindent Guzm\'an R., Lucey J. R., 1993, MNRAS, 263, 47\hfil\break
\noindent Huchra J. P., Geller M. J., 1982, ApJ, 257, 423 (HG)\hfil\break
\noindent Jenkins C. R., 1982, MNRAS, 200, 705\hfil\break
\noindent Kauffmann G., 1996, MNRAS, 281, 487\hfil\break
\noindent Kleinmann S. G., Hall D. N. B., 1986, ApJS, 62, 501\hfil\break
\noindent Larson R. B., Tinsley B. M., Caldwell C. N., 1980, ApJ, 237,
692\hfil\break
\noindent Lucey J. R., Guzm\'an R., Steel J., Carter D., 1997,
MNRAS, 287, 899\hfil\break
\noindent Mobasher B., James P. A., 1996, MNRAS, 280, 895 (paper I)\hfil\break
\noindent O'Connell R. W., 1980, ApJ, 236, 430\hfil\break
\noindent Olsen E. T., 1970, AJ, 75, 764\hfil\break
\noindent Origlia L., Moorwood A. F. M., Oliva E., 1993, A\&A, 280,
536\hfil\break
\noindent Puxley P. J., Doyon R., Ward M. J., 1997, ApJ, 476,
120 (PDW)\hfil\break
\noindent Rose J. A., Bower R. G., Caldwell N., Ellis R. S., Sharples
R. M., Teague P., 1994, AJ, 108, 2054\hfil\break 
\noindent Schweizer F., Seitzer P., 1992, AJ, 104, 1039\hfil\break
\noindent Silva D. R., Bothun G. D., 1998, AJ, 116, 85\hfil\break
\noindent Tinsley B.M., Gunn J.E., 1976, ApJ, 203, 52\hfil\break
\noindent Trager S.C., Worthey G., Faber S.M., Burstein D., Gonzalez J., 
1998, ApJS, 116, 1\hfil\break
\vfil\break
\begin{table}
\centerline{\bf Table 1}
\begin{tabular}{lcccccccc}\\ 
& & & & & & & &\\
Name	&      V    & CO$_{sp}$ &   EW         &  H$\beta$ & M$_B$  &   $\sigma$    &
Mg$_2$ & Cluster\\
& & & & & & &\\    
IC 5157    &  4443  &   0.231 &	2.59  & - & -20.72 &   --	    &
-- & Isol\\
NGC 1052   &  1474  &   0.246 & 3.77  & - & -20.12 &  2.313   &
0.316 & HG44\\
NGC 1600   &  4718  &   0.285 &	4.21  & 1.30 & -22.16 &  2.506   &
0.324 & N1600 gr\\
NGC 6020   &  4397  &   0.241 &	2.77  & 1.70 & -20.15 &  2.296   &
0.306 & Isol\\
NGC 6051   &  9588  &   0.274 &	3.69  & 2.03 & -21.58 &  2.386   &
0.338 & --\\
NGC 6127   &  4609  &   0.238 &	2.88  & 1.33 & -21.02 &  2.355   &
0.316 & Isol\\
NGC 6137   &  9306  &   0.248 &	2.99  & 0.63 & -22.07 &  2.496   &   0.294
& --\\
NGC 6577   &  5205  &   0.227 &	2.86  & - & -20.51 &  --	    &
-- & --\\
NGC 6702   &  4712  &   0.259 &	3.60  & 2.43 & -20.95 &  2.259   &
0.272 & --\\
NGC 6703   &  2365  &   0.257 &	3.71  & 1.74 & -20.52 &  2.233   &
0.280 & --\\
NGC 7385   &  7809  &   0.250 &	3.98  & 1.29 & -21.83 &  2.414   &
0.324 & --\\
NGC 7391   &  3045  &   0.221 &	2.86  & 1.74 & -20.08 &  2.436   &
0.317 & Isol\\
NGC 7454   &  1991  &   0.275 &	3.79  & 2.11 & -19.42 &  2.050   &
0.206 & GH163\\
NGC 7562   &  3636  &   0.275 &	3.66  & 1.74 & -21.05 &  2.385   &
0.291 & GH166\\
NGC 7626   &  3423  &   0.289 &	3.90  & 1.27 & -21.24 &  2.369   &
0.336 & GH166\\
NGC 7680   &  5177  &   0.227 &	2.79  & - & -21.27 &  --	    &
--    & --\\
NGC 7785   &  3849  &   0.231 &	2.90  & 1.66 & -21.14 &  2.464   &
0.296 & Isol\\
E462-G015  &  5827  &   0.234 &	2.75  & - & -21.91 &  2.467   &
0.292 & Isol\\
E468-G013  &  6890  &   0.230 &	2.59  & - & -20.48 &  --	    &
--    & --\\
NGC 375    &  6011  &   0.210 &	3.28  & - & -19.60 &  --	    &
--    & Pisces\\
NGC 380    &  4384  &   0.267 &	3.54  & 0.75 & -20.46 &  2.443   &
0.338 & Pisces\\
NGC 382    &  5217  &   0.259 &	3.36  & 1.06 & -20.25 &  2.185   &
0.269 & Pisces\\
NGC 384    &  4398  &   0.213 &	2.99  & - & -20.02 &  --	    &
--    & Pisces\\
NGC 386    &  5555  &   0.249 &	3.62  & - & -19.49 &  1.785   &
0.232 & Pisces\\
NGC 388    &  5114  &   0.197 &	3.04  & - & -19.29 &  --	    &
--    & Pisces\\
NGC 410    &  5296  &   0.248 &	3.45  & 2.32 & -22.01 &  2.507   &
0.336 & Pisces\\
NGC 6158   &  8914  &   0.233 &	3.15  & 2.03 & -20.70 &  2.280   &
0.277 & A2199\\
NGC 6166   &  9414  &   0.277 &	3.23  & 0.44 & -22.67 &  2.475   &
0.319 & A2199\\
NGC 6173   &  8771  &   0.244 &	3.62  & - & -22.17 &  2.417   &
0.332 & A2197\\
A2199-103  &  9400  &   0.275 &	3.92  & - & -19.32 &  2.252   &
0.300 & A2199\\
A2199-105  &  8677  &   0.245 &	2.66  & - & -19.27 &  2.243   &
0.300 & A2199\\
A2199-121  &  8780  &   0.288 &	3.25  & - & -20.15 &  2.240   &
0.287 & A2199\\
A2199-BO5  &  8704  &   0.277 &	3.58  & - & -20.50 &  2.312   &
0.284 & A2199\\
A2199-BO8  &  9280  &   0.242 &	2.86  & - & -20.24 &  2.146   &
0.269 & A2199\\
A2199-BO26 &  9136  &   0.248 &	3.28  & - & -20.06 &  2.310   &
0.275 & A2199\\
NGC 7720   &  9130  &   0.245 &	3.60  & 0.71 & -22.33 &  2.520   &
0.332 & A2634\\
A2634-BO5  &  9345  &   0.257 &	3.86  & - & -20.69 &  2.450   &
0.315 & A2634\\
A2634-BO4  &  9981  &   0.229 &	2.81  & - & -21.09 &  2.326   &
0.296 & A2634\\
NGC 7728   &  9498  &   0.244 &	3.34  & - & -21.71 &  2.519   &
0.333 & A2634\\
UGC 12733  & 11850  &   0.225 &	2.97  & - & -21.54 &  2.422   &
0.249 & A2634\\
A2634-102  &  9251  &   0.266 &	3.58  & - & -20.14 &  2.291   &
0.265 & A2634\\
A2634-121  &  9586  &   0.209 &	3.38  & - & -18.52 &  2.263   &
0.284 & A2634\\
A2634-124  &  9846  &   0.217 &	2.68  & - & -21.42 &  2.273   &
0.297 & A2634\\
A2634-134  &  9297  &   0.246 &	3.01  & - & -20.04 &  2.345   &
0.301 & A2634\\
A2634-139  &  9520  &   0.245 &	3.81  & - & -20.20 &  2.335   &
0.324 & A2634\\
A2634-1222 &  8088  &   0.229 &	2.79  & - & -20.12 &  2.312   &
0.295 & A2634\\
A2634-1482 &  9347  &   0.252 &	3.25  & - & -21.22 &  2.383   &
0.276 & A2634\\
A2634-BO9  &  9970  &   0.253 &	3.45  & - & -21.54 &  2.317   &
0.291 & A2634\\
A2634-BO13 &  9542  &   0.240 &	3.25  & - & -20.32 &  2.201   &
0.249 & A2634\\
A2634-BO16 &  9337  &   0.250 &	3.38  & - & -20.05 &  2.351   &
0.325 & A2634\\
\end{tabular}
\end{table}

\vfil\break
 
\centerline{\bf Figure Captions}

\noindent {\bf Figure 1.} A histogram showing the distribution of CO EW
values for cluster (solid line) and field (dashed line \& shading) ellipticals.

\noindent {\bf Figure 2.} CO EW as a function of projected distance from
the cluster centre, for 30 ellipticals in rich clusters.

\noindent {\bf Figure 3.} CO EW as a function of the metallicity index
$Mg_2$ for the 44 elliptical galaxies.  Plotted symbols are the same as
for Fig. 2.

\noindent {\bf Figure 4.} Metallicity index $Mg_2$ as a function of total
B-band absolute magnitude for 44 elliptical galaxies. Plotted symbols are the same as
for Fig. 2.

\noindent {\bf Figure 5.} CO EW as a function of total B-band absolute
magnitude for 50 elliptical galaxies. Plotted symbols are the same as
for Fig. 2.

\noindent {\bf Figure 6.} CO EW as a function of H$\beta$ EW
for 19 elliptical galaxies. Plotted symbols are the same as
for Fig. 2.

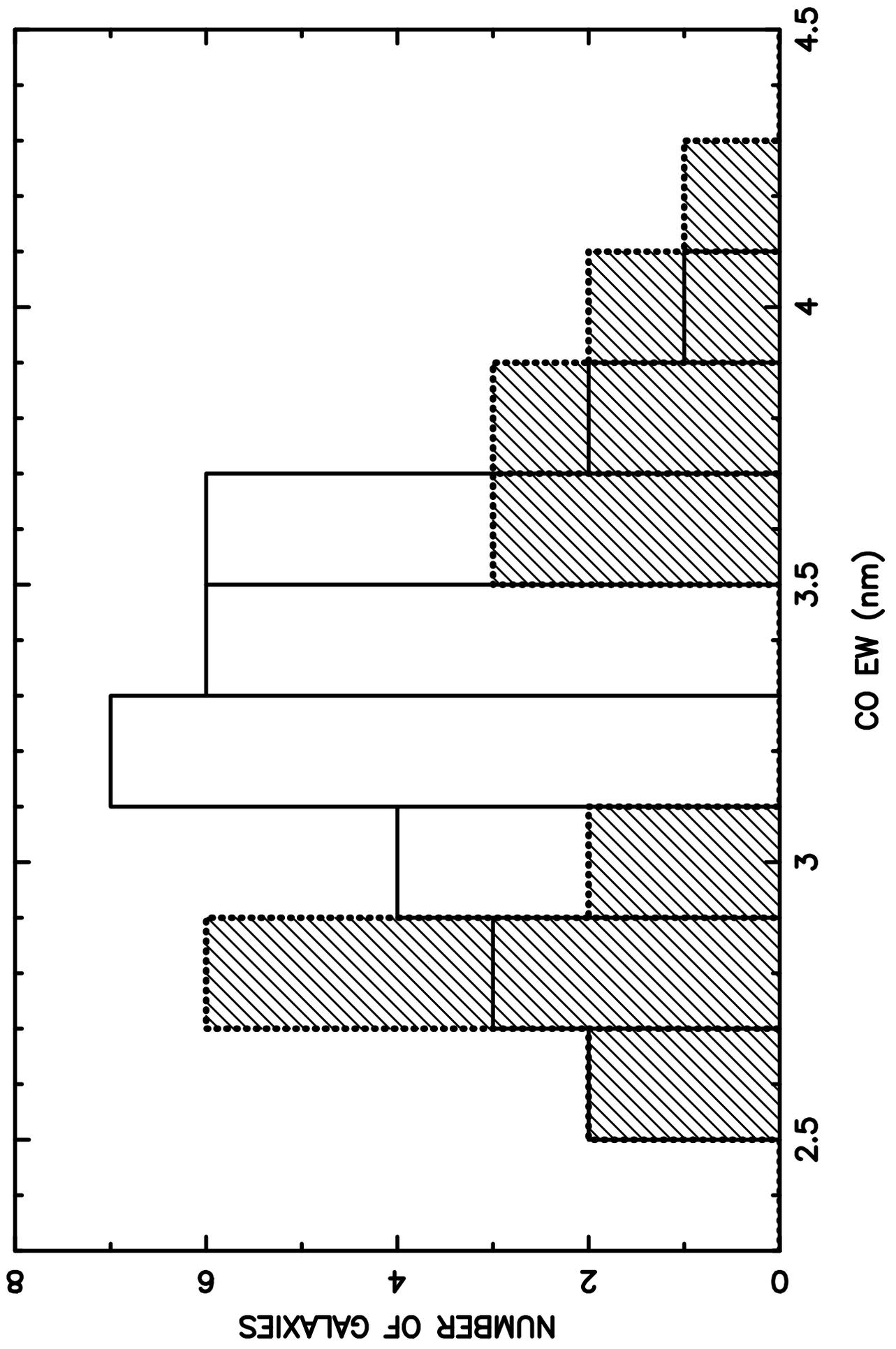
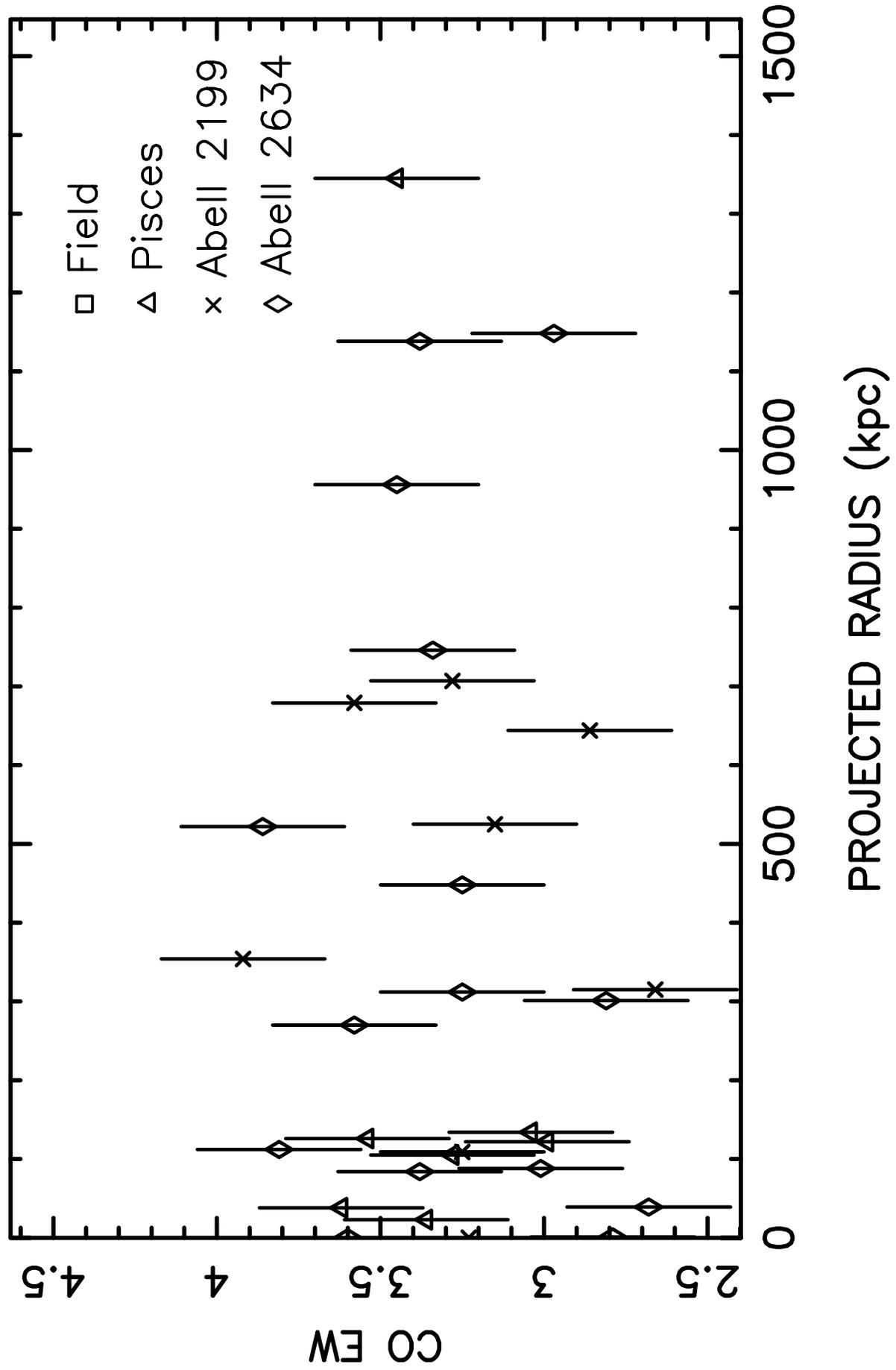
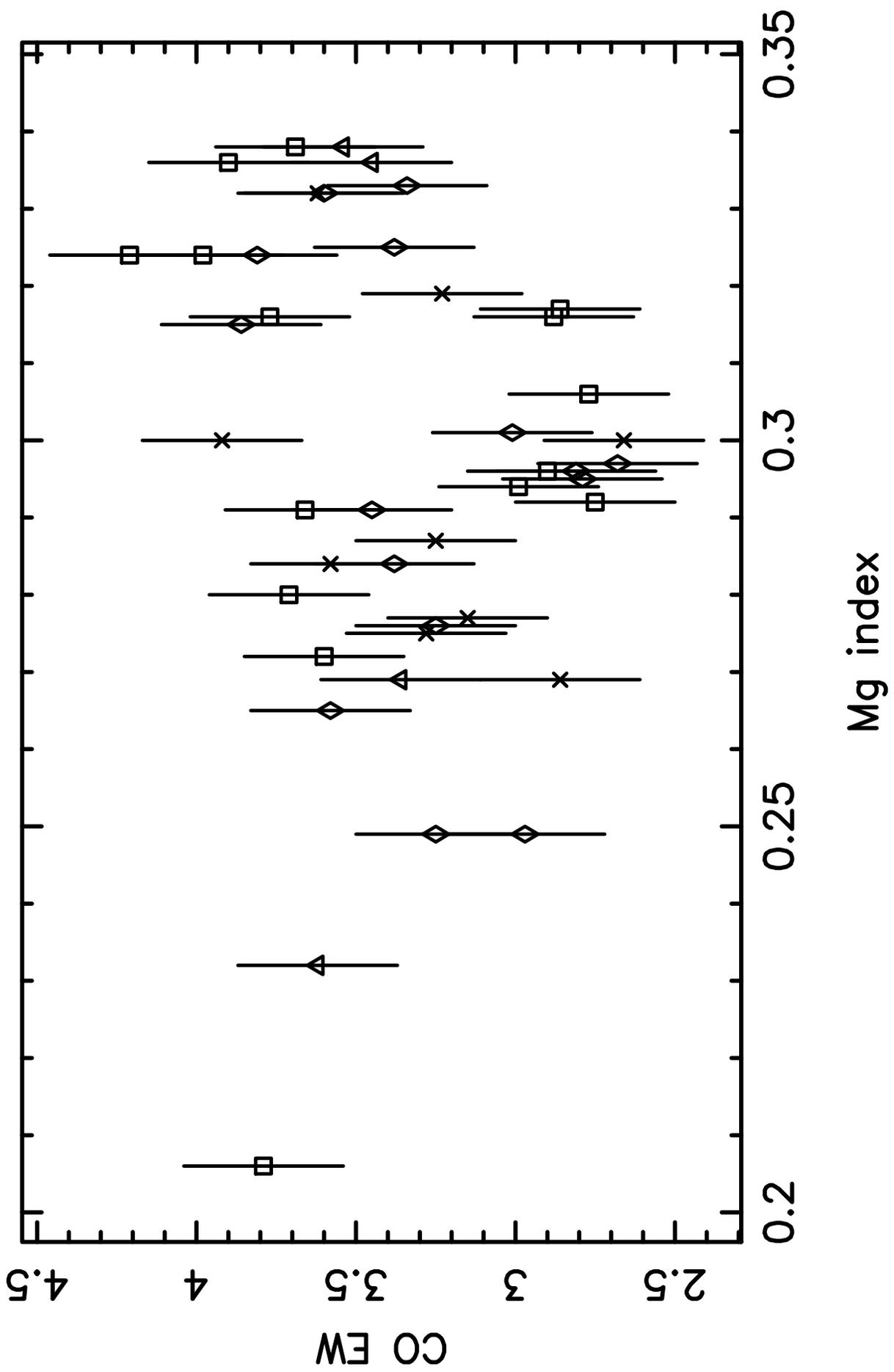
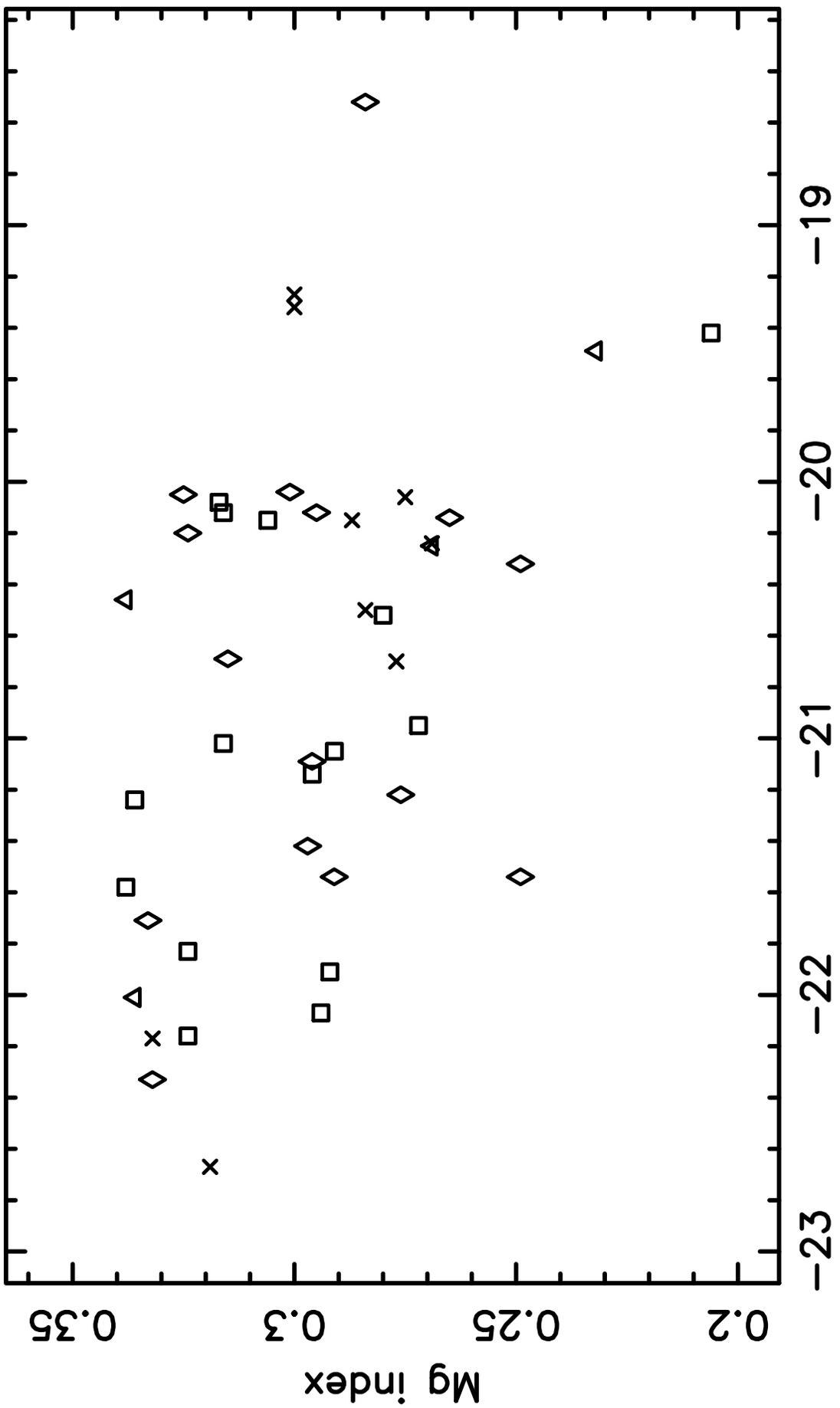
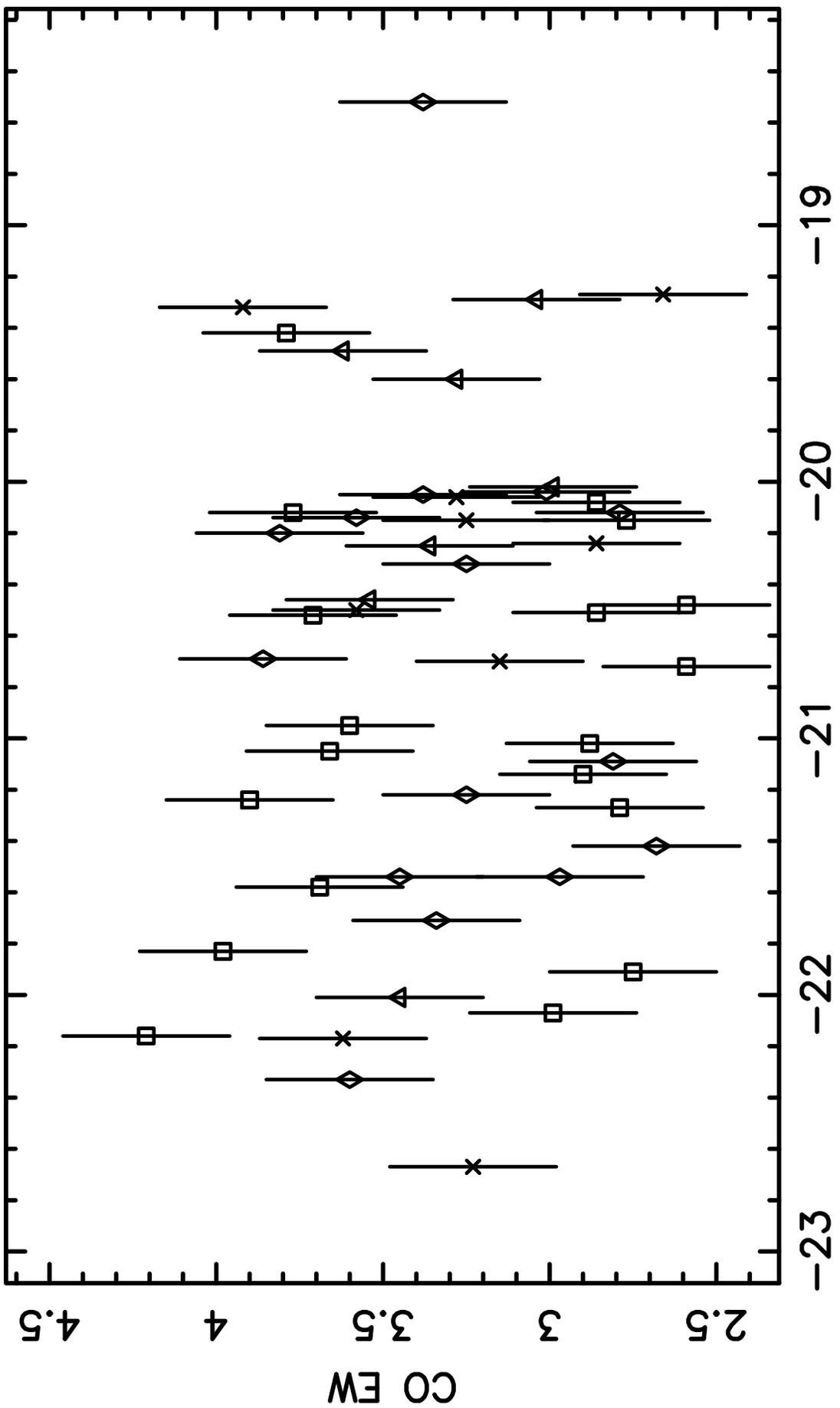
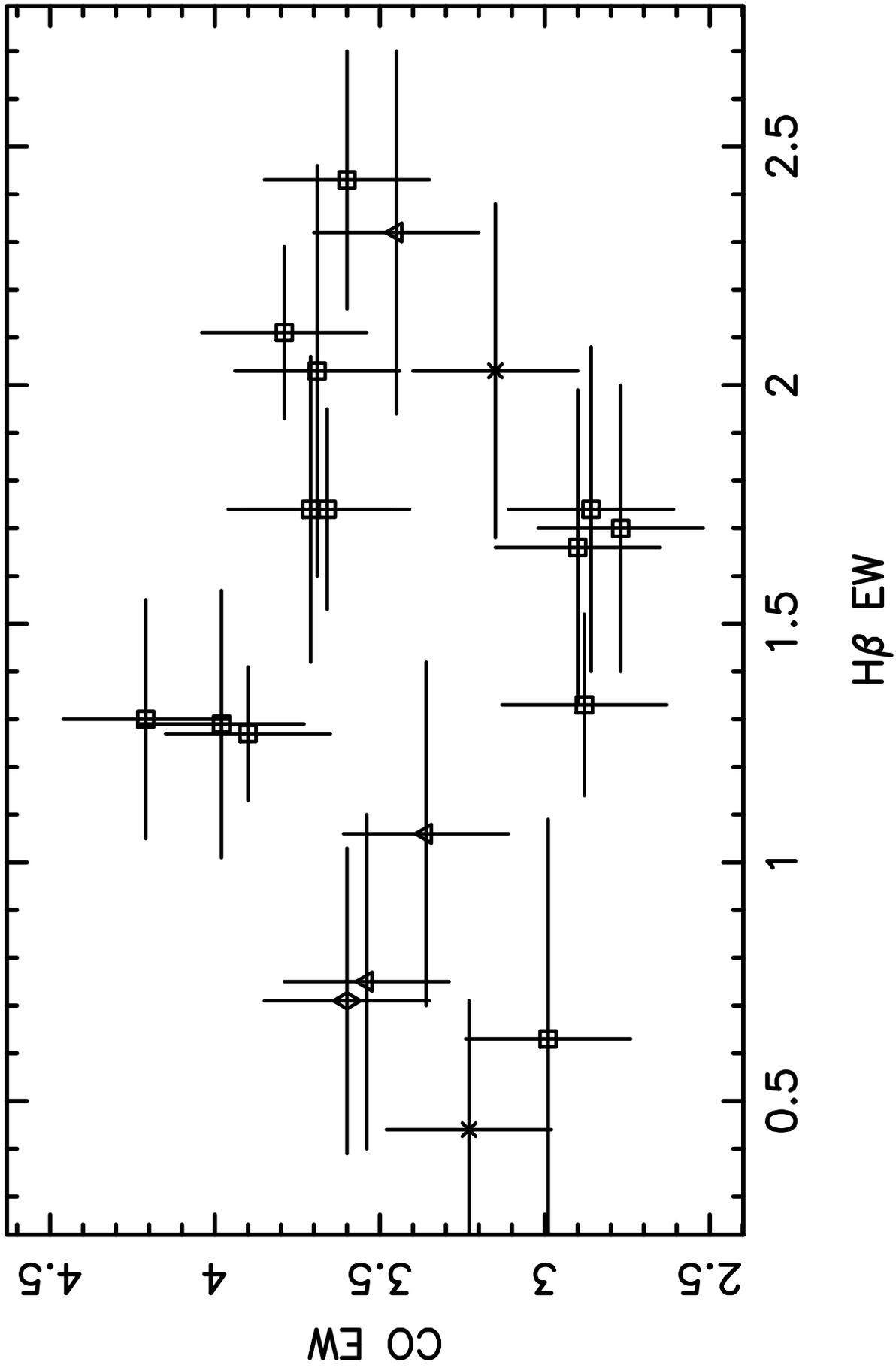
\end{document}